\documentstyle[12pt] {article}

\def\lsim{\compoundrel<\over\sim}
\def\compoundrel#1\over#2{\mathpalette\compoundreL{{#1}\over{#2}}}
\def\compoundreL#1#2{\compoundREL#1#2}
\def\compoundREL#1#2\over#3{\mathrel
     {\vcenter{\hbox{$\buildrel{#1#2}\over{#1#3}$}}}}

\begin{document}

\title{Steering and ro-vibrational effects in the dissociative adsorption and
       associative desorption of H$_2$/Pd(100)}

\author{Axel Gross and Matthias Scheffler\\
Fritz-Haber-Institut, Faradayweg 4-6, 
D-14195 Berlin-Dahlem, Germany} 

\date{}

\maketitle

\vspace{-.7cm}

\begin{abstract}

The interaction of hydrogen with many transition metal surfaces 
is characterized by a coexistence of activated 
with non-activated paths to adsorption  with a broad distribution 
of barrier heights. By performing six-dimensional quantum dynamical 
calculations  using a potential energy surface derived 
from {\it ab initio} calculations for the system H$_2$/Pd(100)
we show that these features of the potential energy surface
lead to strong steering effects in the dissociative adsorption and
associative desorption dynamics.
In particular, we focus on the coupling of the translational, rotational
and vibrational degrees of freedom of the hydrogen molecule in the
reaction dynamics. 

\end{abstract}


\section{Introduction}

It is a long-term goal in surface science to understand
catalytic reactions occuring at surfaces. Obviously, the single steps
of these often rather complicated processes are more effectively
studied at simple systems. In particular, the dissociative 
adsorption and associative desorption of hydrogen on metal surfaces 
has served as a benchmark system, both experimentally and theoretically
(see, e.g., Refs.~\cite{Ren94,Hol94,Dar95} and references therein). 
Since the mass mismatch between hydrogen and a metal substrate 
is rather large, the crucial process in the dissociative adsorption
for these particular systems is the conversion of 
translational and internal energy of the hydrogen molecule into 
translational and vibrational energy of the adsorbed hydrogen atoms. 
If in addition no surface rearrangement occurs upon adsorption,
the substrate degrees of freedom can be neglected and the dissociation
dynamics can be described in terms of potential energy surfaces (PES)
which take only the molecular degrees of freedom into account.

\vspace{-17.4cm}

{\tt subm.~to the Proceedings of ISSP-18, June 1996, Poland, to appear in
Prog.~Surf.~Sci.}

\vspace{17.cm}

The PES for the dissociative adsorption of a diatomic molecule 
neglecting the substrate degrees of freedom is still six-dimensional.
These PESs now become available by elaborate density-functional
calculations \cite{Ham94,Whi94,Wil95,Wil96,Wie96}.
However, in order to understand the reaction dynamics one has to
perform dynamical calculations on these potentials.
Because of its light mass hydrogen has to be described quantum
mechanically. Only recently it has become possible to
perform dynamical calculations of the dissociative adsorption
and associative desorption where {\em all} degrees of freedom of
the hydrogen molecule are treated quantum mechanically \cite{Gro95}. 
These calculations have established the importance of high-dimensional
effects in the reaction dynamics.

For example, molecular beam experiments of the 
dissociative adsorption of H$_2$ on various
transition metal surfaces like Pd(100) \cite{Ren89}, Pd(111) and Pd(110)
\cite{Res94}, W(111) \cite{Ber92}, W(100) \cite{Ber92,But94,Aln89},
W(100)--c(2$\times$2)Cu \cite{But95} and Pt(100) \cite{Dix94}  
revealed that the sticking probability in these systems
initially decreases with increasing kinetic energy of the beam.
High-dimensional quantum dynamical calculations have shown that
steering effects can cause such an initial decrease in the
sticking probability \cite{Gro95,Kay95}; it is not necessarily
due to a precursor mechanism, as was widely believed. 
In this paper we briefly review
the most important aspects of the steering mechanism, but
mainly focus on the combined influence of the translational,
rotational and vibrational degrees of freedom of the hydrogen molecule
on the adsorption and desorption process. In the next section
the theoretical background will be introduced before the results
of the dynamical calculations will be discussed.

\section{Theoretical background}

The potential energy surface of H$_2$/Pd(100)
has been determined using density-functional
theory together with the generalized gradient approximation (GGA) \cite{Per92} 
and the full-potential linear augmented plane wave method \cite{Bla93,Koh95}. 
{\em Ab initio} total energies have been evaluated for more than 250 
configurations \cite{Wil95} and have been parametrized in a suitable form for the 
dynamical calculations \cite{Gro95}.

\begin{figure}[t]
\unitlength1.cm
\begin{minipage}{7.cm}
   \begin{picture}(11.0,10.5)
      \includegraphics{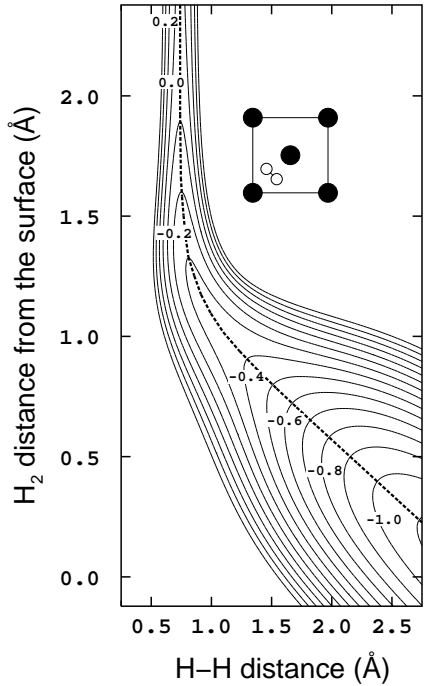}
   \end{picture}
\end{minipage}
\hspace{1.cm}
\begin{minipage}{7.cm}
   \caption{Contour plot of the PES along a
            two-dimensional cut through the
            six-dimensional coordinate space of H$_2$/Pd\,(100).
            The inset shows the
            orientation of the molecular axis and the lateral
            H$_2$ center-of-mass coordinates. The coordinates 
            in the figure are the H$_2$ center-of-mass distance 
            from the surface $Z$ and the H-H interatomic distance $d$. The 
            dashed line is the optimum reaction path.
            Energies are in eV per H$_2$ molecule.
            The contour spacing is 0.1~eV.  }
\label{elbow}
\end{minipage}
\end{figure}

Figure \ref{elbow} shows a cut through
the PES of H$_2$/Pd\,(100), where the most 
favourable path towards dissociative adsorption is marked by the dashed line. 
For this path there is no energy barrier hindering
dissociation, i.e., the adsorption is non-activated.
However, the majority of pathways towards dissociative adsorption
has in fact energy barriers with a rather broad
distribution of heights and positions, as the detailed total-energy 
calculations showed~\cite{Wil95}, i.e. the PES is strongly anisotropic 
and corrugated. This has important consequences, as will be shown below.

The quantum dynamics is determined in a coupled-channel
scheme within the concept of the {\em local reflection matrix} (LORE) 
\cite{Bre93}. This numerically very stable method
is closely related to the logarithmic derivative of the solution matrix
and thus avoids exponentially increasing outgoing waves which could
cause numerical instabilities. 
The reported calculations, which take all degrees of freedom of the hydrogen
molecule into account, are still only possible if all symmetries of
the scattering problem are utilized.

\section{Results}

\begin{figure}[t]
\unitlength1cm
\begin{center}
   \begin{picture}(10,8.5)
   \includegraphics{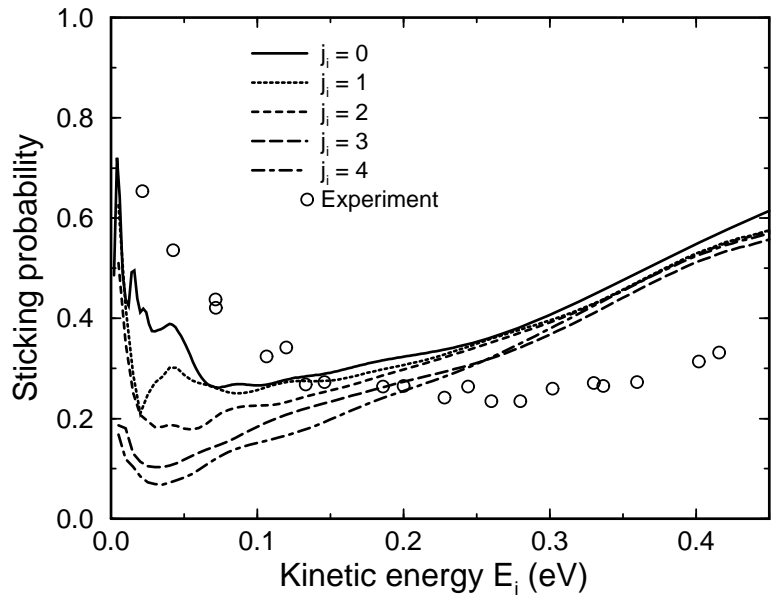}
   \end{picture}
\end{center}
   \caption{Sticking probability versus kinetic energy for
            a H$_2$ beam under normal incidence on a Pd(100) surface.
            Theory: orientationally averaged sticking probability  
            for different initial rotational quantum 
            numbers~$0 \le j_i \le 4$ of the incoming molecular beam. 
            The molecular beams are assumed to have an energy spread of
            $\Delta E / E_i = 2 \Delta v / v_i = 0.2$ \protect{\cite{Ren89}} 
            ($E_i$ and $v_i$ are the initial kinetic energy and velocity, 
            respectively).
            Experimental data (circles) are from ref.~\protect{\cite{Ren89}}.} 
\label{stickrot}
\end{figure}

Figure \ref{stickrot} presents six-dimensional quantum dynamical
calculations of the sticking probability  as a 
function of the kinetic energy of a H$_2$ beam under normal incidence
on a Pd(100) surface for different initial rotational states
averaged over the azimuthal quantum numbers. 
In addition, the results of a molecular beam experiment are shown \cite{Ren89}.
Quantum mechanically determined sticking probabilities 
for hydrogen at surfaces with an attractive well exhibit an oscillatory
structure as a function of the incident energy~\cite{Gro95,Kay95,Dar90,Gro95b},
reflecting the opening of new scattering channels and 
resonances \cite{Dar90,Gro95b}. These structures are known for a long time
in He and H$_2$ scattering \cite{stern} and also in LEED \cite{LEED}. For 
H$_2$/Pd(100), however, measuring these oscillations is a very demanding
task. They are very sensitive to surface imperfections like adatoms
or steps \cite{Gro96c}. 
Since we do not focus on these oscillations here, for the results of
Fig.~\ref{stickrot} we have assumed a velocity spread of the incoming
beam typical for the experiment \cite{Ren89} so that the oscillations
are smoothed out.

The initial decrease of the sticking probability found in the
experiment is well-produced for molecules initially in the 
rotational ground state $j_i = 0$. The high sticking probability
at low kinetic energies is caused by the steering effect: Slow
molecules can very efficiently be steered to non-activated pathways
towards dissociative adsorption by the attractive forces of the
potential. This mechanism becomes less effective at higher 
kinetic energies where the molecules are too fast to be focused
into favourable configurations towards dissociative adsorption.
This causes the initial decrease of the sticking probability. If the
kinetic energy is further increased, the molecules will eventually
have enough energy to directly traverse the barrier region leading
to the final rise in the sticking probability.

The dynamical origin of the steering effect is reflected in the
dependence of the sticking probability on the initial rotational
state of the impinging molecules. Rapidly rotating molecules
will rotate out of favourable configurations towards dissociative
adsorption, i.e. the steering mechanism will be suppressed.
This rotational hindering is demonstrated in Fig.~\ref{stickrot}:
at low kinetic energies the sticking probability decreases with
increasing initial rotational quantum number $j_i$. At higher
kinetic energies, where direct dissociative adsorption becomes
dominant, rotational hindering is less important. As a consequence
in this energy range the sticking probability is almost independent 
of the initial rotational state.

\begin{figure}[tb]
\unitlength1cm
\begin{center}
   \begin{picture}(10.,8.5)
      \includegraphics{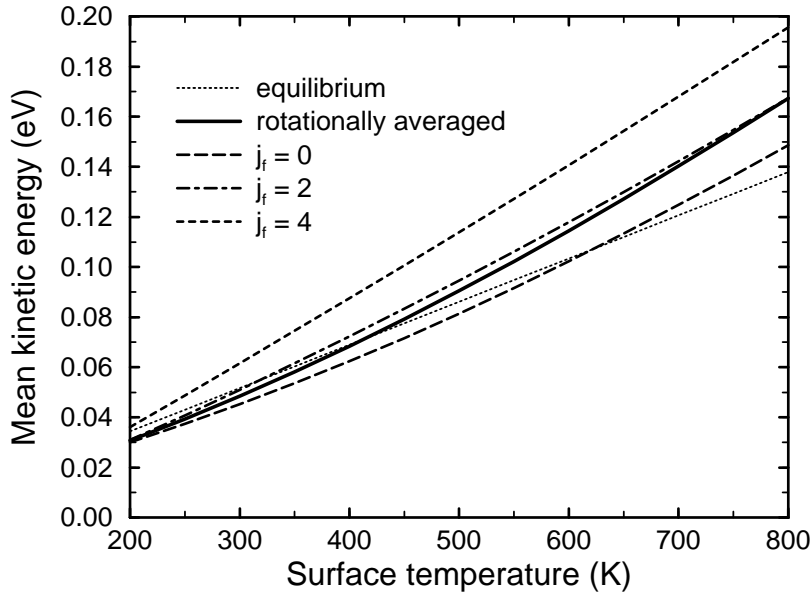}
   \end{picture}

\end{center}
   \caption{Mean kinetic energy of hydrogen molecules desorbing
       perpendicurlarly from a Pd(100) surface for different
       final rotational states as a function of the surface
       temperature. 
       Thin dotted line: result for molecules in equilibrium
       with the surface temperature;
       thick solid line: summed over all final rotational states;
       dashed and dash-dotted lines: results for final
       rotational states $j_f =$~0, 2, 4.}

\label{kindes}
\end{figure}

The predicted rotational hindering in the sticking probability
at low kinetic energies \cite{Gro96} has in the meantime 
been confirmed by experiment \cite{Beu95}. The influence of the
steering mechanism on the dissociation dynamics can also be
demonstrated at the time-reverse process, namely the associative
desorption. Invoking the principle of microscopic reversibility,
any degree of freedom that hinders dissociative adsorption
will in thermal associative desorption be less populated
than expected for thermal equilibrium with the surface
temperature $T_s$. At low kinetic energies increasing the initial rotational 
state as well as increasing the kinetic energy hinders dissociative
adsorption. In desorption this leads to the so-called rotational
cooling, which was found experimentally \cite{Sch91} and confirmed
theoretically \cite{Gro95}, as well as to translational cooling.
The translational cooling is demonstrated  in Fig.~\ref{kindes}
where the mean kinetic energy in desorption as a function of the
surface temperature for different final rotational states is shown.
The thin dotted line corresponds to thermal equilibrium with the
surface temperature. At low surface temperatures the mean kinetic
energy of molecules desorbing in rotational states $j_f \le 2$ is
indeed smaller than the equilibrium value. Translational cooling
is also found if the  mean kinetic energy is summed over all 
rotational states at low surface temperatures since at these 
temperatures the desorption flux is mostly populated by molecules
in low rotational states. For large rotational
quantum numbers the steering effect in adsorption is suppressed 
(see Fig.~\ref{stickrot}), consequently for $j_f = 4$ only
translational heating is found in desorption.
Equivalently, at higher kinetic energies the sticking probability 
rises regardless of the rotational state which leads to 
translational heating at higher surface temperatures 
for all final rotational states.

However, Fig.~\ref{kindes} shows that the mean kinetic energy in 
desorption does not differ significantly from the thermal equilibrium value.
State-resolved measurements of the mean kinetic energy in desorption 
were done in the surface temperature range 450~K~$\le T_s \le$~850~K
\cite{Sch92b}. For H$_2$ desorption the experiments do not show
any translational heating, within the error bars the experiments are, however,
consistent with our results. For D$_2$ desorption on the other hand, 
translational heating has been found \cite{Sch92b} in good agreement
with our H$_2$ calculations.

\begin{figure}[t]
\unitlength1cm
\begin{center}
   \begin{picture}(10,8.5)
   \includegraphics{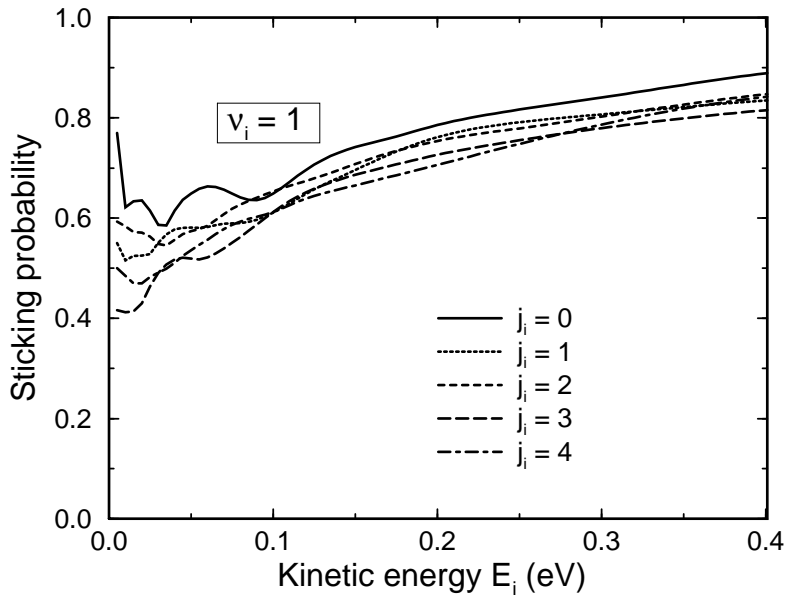}
   \end{picture}
\end{center}
   \caption{Sticking probability versus kinetic energy for
            a H$_2$ beam under normal incidence on a Pd(100) surface.
            Orientationally averaged sticking probability  
            for different initial rotational quantum 
            numbers~$0 \le j_i \le 4$ for molecules initially in
            the first excited vibrationally state $\nu_i = 1$. 
            The molecular beams are assumed to have an energy spread of
            $\Delta E / E_i = 2 \Delta v / v_i = 0.2$. } 
\label{stickv1rot}
\end{figure}

Interestingly, strong vibrational heating in D$_2$ desorption 
from Pd(100) has been found experimentally~\cite{Sch89}. Usually
this is associated with a so-called late barrier to adsorption
as found in the system H$_2$/Cu \cite{Ham94,Whi94,Wie96}. Since the
adsorption of H$_2$/Pd(100) is apparently non-activated
\cite{Wil95,Ren89,Sch92b}, the origin of the vibrational heating
has been controversely debated~\cite{Sch92,Bre94,Dar92b,Dar93}.
Just recently we have shown that the vibrational effects in the
system are caused by the lowering of the H-H vibrational 
frequency during the dissociation and the multi-dimensionality of the 
potential energy surface \cite{Gro96b}. 

From Fig.~\ref{elbow} it
is apparent that the H-H vibrational frequency is strongly lowered
during the dissociation. This is also true for other pathways
to dissociative adsorption. A detailed \mbox{analysis} of the dynamics
has further shown \cite{Gro96b} that the H-H vibrations follow
the change of the vibrational frequency almost adiabatically
during the dissociation. This leads to an effective 
vibrational--translational energy transfer, or equivalently, to
an effective lowering of the PES while the molecules approach the
surface. The crucial point is that this
lowering depends on the vibrational state: the higher the vibrational
quantum number is, the larger the effective lowering is. Thus molecules
in the first excited vibrational state experience a
lower barrier range than molecules in the vibrational ground state.

In fact, for molecules in the first excited vibrational state the PES
due to the lowering is so attractive that the steering
mechanism is hardly operative. This is demonstrated in Fig.~\ref{stickv1rot}
where the sticking probability for molecules initially in the first
excited vibrational state $\nu_i =1$ for different initial rotational 
states~$j_i$ is shown. For $j_i = 0$ only at very low kinetic energies
\mbox{$E_i \lsim 0.02$~eV} there is a significant decrease in the sticking
probability with increasing kinetic energy. \mbox{At $E_i = 0.1$~eV}
the sticking probability for vibrationally excited molecules is about
2.5~times larger than for molecules in the vibrational ground state
(compare Fig.~\ref{stickrot}). In desorption this leads to a population
of the first excited vibrational state which is two times larger than
expected for thermal equilibrium with the surface temperature at
$T_s = $~700~K \cite{Gro96b}.

The fact that for vibrationally excited molecules already at small
kinetic energies direct dissociative adsorption is the dominant
mechanism is also reflected by the dependence of the sticking probability
on the initial rotational state for vibrationally excited molecules.
Fig.~\ref{stickv1rot} shows that this dependence is much less 
pronounced than for molecules in the vibrational ground state. 
Due to the effectively lowered potential vibrationally excited
molecules are so fast that the rotational hindering does not play
any crucial role.

So far all the results presented were averaged over the azimuthal
quantum number $m$, i.e, the results corresponded to orientationally
averaged properties. Now the H$_2$/Pd(100) PES is strongly anisotropic
with regard to the molecular orientation. The most favourable
configuration towards dissociative adsorption is with the molecular
axis parallel to the surface. Molecules that hit the surface in an
upright position cannot dissociate, they are reflected back into the 
gas-phase. It is true that quantum mechanics does not allow for
non-rotating, oriented molecules in the gas-phase, however, rotating
molecules can show a preferential orientation. Molecules with 
azimuthal quantum number~$m = j$ have their axis preferentially 
oriented parallel to the surface. It has been shown theoretically that
these molecules rotating in the so-called helicopter fashion  
dissociate more easily than molecules rotating in the cartwheel fashion 
($m = 0$) with their rotational axis preferentially parallel to the
surface \cite{Gro95}.

Experimentally it is hard to align a molecular beam of hydrogen.
Again one can study the time-reverse process, the associative desorption.
By laser-induced fluorescence (LIF) it is possible to measure
the rotational alignment parameter $A_0^{(2)}(j)$ \cite{Gre83}, 
which is given by
\begin{equation}
A_0^{(2)} (j) \ = \ \left\langle \frac{3J_z^2 \ - \ {\bf J}^2}{{\bf J}^2}
\right\rangle_j 
\end{equation}
$A_0^{(2)}(j)$ corresponds to the quadrupole moment of the orientational
distribution and assumes values of $-1 \ \leq \ A_0^{(2)} (j) \ \leq 2$.
Molecules rotating preferentially in the cartwheel fashion have an
alignment parameter $A_0^{(2)} (j) < 0$, for molecules rotating 
preferentially in the helicopter fashion $A_0^{(2)} (j) > 0$.

\begin{figure}[tb]
\unitlength1cm
\begin{center}
   \begin{picture}(10.,8.5)
      \includegraphics{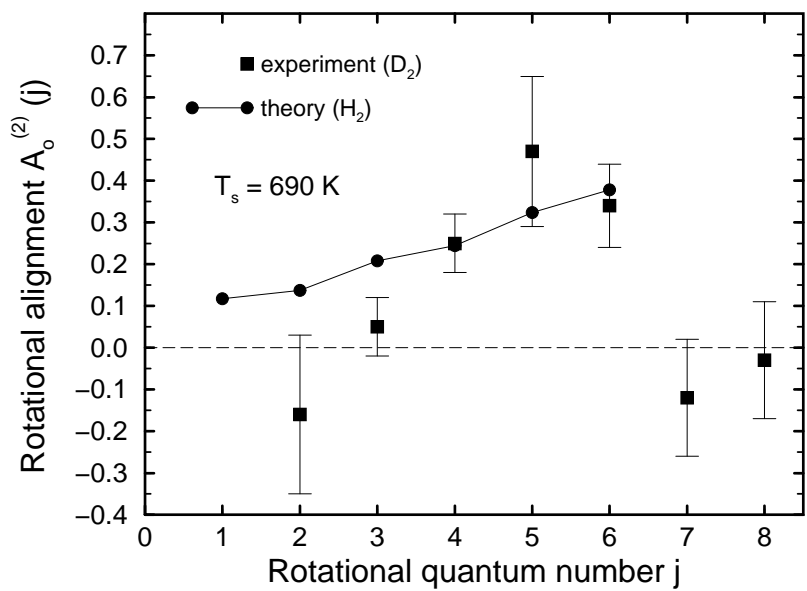}
   \end{picture}

\end{center}
   \caption{Rotationally alignment of hydrogen molecules desorbing
       from a Pd(100) surface. 
       Boxes: experimental results for D$_2$ \protect{\cite{Wet96}}. 
       Circles: 6-D calculations for H$_2$.}

\label{rotaligndes}
\end{figure}

Figure~\ref{rotaligndes} shows a comparison between experiment \cite{Wet96}
and theory of the rotational alignment of hydrogen desorbing from a
Pd(100) surface at a surface temperature of $T_s = 690$~K. Note that
the experiments were done for D$_2$, while the calculations were performed
for H$_2$, hence the comparison has to be done with caution. Still
the agreement for $j \le 6$ is quite satisfactory. Indeed the molecules
desorb preferentially with their molecular axis parallel to the surface
thus reflecting the anisotropy of the PES. Due to computational
restrictions the rotational alignment could only be calculated for
$j \le 6$. For $j = 7$ and $j =8$ the experiments show a vanishing
alignment within the error bars which is rather surprising. It was
argued that for these high rotational state an energy transfer from
the rotations to the reaction coordinate could suppress the effect
of the potential anisotropy \cite{Wet96}. Certainly these results 
deserve further clarification.

\section{Conclusions}

In conclusion, we reported a six-dimensional quantum dynamical
study of dissociative adsorption and associative desorption of
H$_2$/Pd\,(100). We have briefly reviewed the steering mechanism,
in particular in the rotational coordinates,
which leads to the initial decrease of the sticking probability
with increasing kinetic energy. Due to the multi-dimensionality
of the relevant potential energy surface the translational, rotational
and vibrational degrees of freedom of the hydrogen molecule are
strongly coupled during the dissociation. Our results establish
the importance of a high-dimensional dynamical treatment in order
to understand reactions at surfaces.

\end{document}